\documentclass{aa}
\usepackage{txfonts}
\usepackage{graphicx}
\usepackage{amsmath}
\usepackage{natbib}
\usepackage{amssymb}
\bibpunct{(}{)}{;}{a}{}{,} 
\begin{document}
\title{Scale invariant cosmology  I: the vacuum  and the cosmological constant }
\titlerunning{Vacuum properties}

\author{Andr\'e Maeder
}
\authorrunning{Maeder}

\institute{Geneva Observatory, Geneva University, CH--1290 Sauverny, Switzerland\\
email: andre.maeder@unige.ch
}

\date{Received  / Accepted }


\abstract
{The source of the acceleration of the expansion of the Universe is still unknown.}
{We examine some consequences of the possible scale invariance
 of the empty space at large scales.}
{ The central hypothesis of this work is that, at macroscopic and large scales where General 
Relativity may be applied, the empty space in the sense it is used in the Minkowski metric, is also scale invariant.}
{ It is shown that if this applies,  the Einstein cosmological constant  $\Lambda_{\mathrm{E}}$ and   the scale factor  $\lambda$ of the scale invariant framework are related by two differential equations. }
{}


\maketitle

 \section{Introduction}

The cause  of the accelerating expansion of the Universe \citep{Riess98,Perl99} is one of the
major scientific problems at present. It leads to many fundamental studies.
These concern  the observational evidences of the acceleration \citep{Shap06,Frie08,Cerv11},
the problems related to the cosmological constant $\Lambda$ \citep{Carr92,Frie08},
 the possibilities of a modified gravity \citep{Milgr83,Milgr09,Wesson83,Wesson15},
the huge discrepancy between the astronomical estimates of $\Lambda$ and the values
 derived from the vacuum energy in particle physics \citep{Wein89,Sola13},
the nature of the so-called dark energy and  the search of possible dark matter candidates
in  astroparticle physics \citep{Feng10,Port11}.
Among the about 50'000 papers with  "cosmological constant" in their title  according to the NASA ADS data basis,
there is an impressive variety of suggestions and tentative
explanations of various natures, as also illustrated by some of the above reviews.
We apologize   for adding some pages to the above mentioned paper avalanche.

The laws of physics generally are not unchanged under a change of scale, a fact originally  discovered by Galileo Galilei  
as recalled by \citet{Feynman63},  who mentions that  Galileo realized {\emph{"that the strengths of materials were not in exactly 
the right proportion to their sizes"}}.  The scale references 
appear closely related  to the material content of the medium.  Even the vacuum
at the quantum level is  not scale invariant, since quantum effects produce some units
of time and length. This point  is as a matter of fact a historical argument, already forwarded long time ago against the theory  of \citet{Weyl23}, who built a geometry that added the scale invariance to the general covariance properties of the gravitational
 field equations. 

  Let us  consider the space at  macroscopic and  large scales, for example such as the scales considered in  cosmology. There,
 the  empty space, in the sense it is used for example in the case of the Minkowski metric, 
does not appear to have  preferred scales of length or
time, even if a particular velocity is considered in Special Relativity.   We now make the hypothesis that the empty space at large scales  is scale invariant. The purpose of this work is to explore some consequences of this particular  hypothesis.  We note that the 
cosmological scales  differ by an enormous factor, up to  ~$10^{39}$, from the nuclear scales where quantum effects intervene.
Thus, in the same way as we may use Newton or Einstein theory at macroscopic and large scales, even if we do not have a quantum theory of gravitation,
we may consider that the large scale empty space is scale invariant, even if this  is
not true at   scales where  quantum effects   intervene.
We remark that it is not uncommon for a physical law to be valid under certain scales or under some  conditions.

 Scale invariance of a system of equations means that the equations do not change
by a transformation of the space and time coordinates  of the form  $ds'  \rightarrow \lambda(x^{\mu}) \, ds$.
There  $\lambda(x^{\mu})$ is the scale factor, where $x^{\mu}$ represents space-time coordinates ($\lambda$ may only
have a time dependence as discussed below).
We  shall  explore some implications  of a possible scale invariance of  the empty space
at  large scales.

A strong reason for doing
that has been emphasized by \citet{Dirac73}: {\emph{It appears as one of the fundamental principles of Nature that the equations
expressing basic laws should be invariant under the widest possible group of transformations.}} It is well known for example that the Maxwell equations of electrodynamics  in absence of charges and currents show the property of scale invariance.
While
scale invariance has often been studied in relation with possible variations of the gravitational constant $G$,
no such hypothesis of variable $G$ is considered here. 
We do not know whether the above hypothesis of scale invariance applies. However, it is by carefully examining the implications of such
a hypothesis that we will  find whether it corresponds to Nature or not.

In Sect. 2, we briefly  recall  the basic scale invariant field equations necessary for the present study.
In Sect. 3, we apply these equations to the  macroscopic space where the Minkowski metric applies and obtain some fundamental relations
between the scale parameter  $\lambda$ (with its derivatives) and Einstein cosmological constant. Section 4 provides the conclusions.

\section{The basic scale invariant field equations}

We make some recalls about scale invariance, limiting them to the necessary minimum. More developments can be found in
works on the scale covariant theory by giants of Physics like \citet{Eddi23}, \citet{Dirac73} and \citet{Canu77}. Their works are  based on some particular case
of  Weyl's geometry \citep{Weyl23}. We recall that General Relativity is not scale invariant. In
its 4-dimensional space, the element interval  $ds' \,^2$ in coordinates
 $x'\,^{\mu} $ writes $ ds'\,^2  \,= \, g'_{\mu \nu} dx'\,^{\mu} \, dx'\,^{\nu}$, (in this work
the symbols with a prime refer to the space of General Relativity).
A scale (or gauge) transformation is considered to a new  coordinate system $x^{\mu}$ with the following relation between the two systems,

\begin{equation}
ds' \, = \, \lambda(x^{\mu}) \, ds \, ,
\label{lambda}
\end{equation}

\noindent
where   $ds^2  =  g_{\mu \nu} dx^{\mu} \, dx^{\nu} $
is the line element in the new  more general framework where scale invariance is supposed to be a fundamental property in addition 
to the general covariance of General Relativity.
Parameter $\lambda(x^{\mu})$ is the scale factor connecting the two line elements. The Cosmological Principle of space  homogeneity and isotropy in cosmology  demands
that the scale factor   only  depends on time.
From the above definitions, we  have  a conformal transformation between the metrics of the two coordinate systems,

\begin{equation}
 g'_{\mu \nu} \, = \, \lambda^2 \, g_{\mu \nu} \, .
\label{gmunu}
\end{equation}

\noindent
In this framework, scalars, vectors or
tensors  that  transform like

\begin{equation}
Y'^{\, \nu   }_{\mu}  \, =  \, \lambda^{n} \, Y^{\nu}_{\mu} \, ,
\label{coco}
\end{equation}

\noindent
 are respectively called coscalars, covectors or  cotensors  of power $n$. If n=0, one has an inscalar, invector or intensor, such objects are
 invariant to the scale transformation   (\ref{lambda}). The term scale covariance  (called co-covariance by Dirac) refers
to the general case of transformations (\ref{coco}) with possibly  different powers $n$, while we reserve the term scale invariance more specifically to the case    $n=0$.

An extensive cotensor analysis has been developed by the above mentioned authors, see also \citet{Bouv78}.  The derivative
 of a  scale invariant object is not in general scale invariant.
Thus,  co-covariant derivatives of the first and second order have been developed preserving  scale covariance.
For example, the co-covariant derivatives  $A_{\mu * \nu}$ and $A^{\mu}_{ * \nu}$ of a co-vector $A_\mu$ become

\begin{eqnarray}
A_{\mu * \nu} \,= \, \partial_\nu A_\mu - ^*\Gamma^{\alpha}_{\mu \nu} A_{\alpha} -n \kappa_\nu A_\mu  \, , \\ [2mm]
A^{\mu}_{ * \nu} \,= \, \partial_\nu A^\mu + ^*\Gamma^{\mu}_{\nu \alpha} A^{\alpha} -n \kappa_\nu A^\mu  \,  \\ [2mm]
\mathrm{with} \quad  ^*\Gamma^{\alpha}_{\mu \nu}=\Gamma^{\alpha}_{\mu \nu} - g^\alpha_\mu \kappa_\nu - g^\alpha_\nu \kappa_\mu
 + g_{\mu \nu}\kappa^\alpha \, .
\end{eqnarray}

\noindent
There, $^*\Gamma^{\alpha}_{\mu \nu}$ is a modified Christoffel symbol, and
$\Gamma^{\alpha}_{\mu \nu}$ is the usual Christoffel symbol. The term
$\kappa_{\mu}$ is called the coefficient of metrical connection, it is
\begin{equation}
\kappa_{\nu} \, = \, -\frac{\partial}{\partial x^{\nu}  } \, \ln \lambda \, .
\label{kappa}
\end{equation}

\noindent
In the scale covariant theory, it is as a fundamental quantity as are the $g_{\mu \nu}$.
A modified  Riemann-Christoffel tensor  $R^{\nu}_{\mu \lambda \rho}$,  its contracted form  $R^{\nu}_{\mu}$ and the total
curvature $R$ also have their
corresponding terms  \citep{Eddi23,Dirac73,Canu77}. The last
two are

\begin{equation}
R^{\nu}_{\mu} = R'^{\nu}_{\mu}  - \kappa^{; \nu}_{\mu}  - \kappa^{ \nu}_{;\mu}
 - g^{\nu}_{\mu}\kappa^{ \alpha}_{;\alpha}  -2 \kappa_{\mu} \kappa^{\nu}
+ 2 g^{\nu}_{\mu}\kappa^{ \alpha} \kappa_{ \alpha}  \, ,
\label{RC}
\end{equation}
\begin{equation}
R \, = \, R' -6 \kappa^{\alpha}_{; \alpha}+6 \kappa^{\alpha} \kappa_{\alpha} \, .
\label{RRR}
\end{equation}

\noindent
There,  $R'^{\nu}_{\mu}$ and $R'$ are the usual expressions
in General Relativity.
The symbol ``;`` indicates a derivative.
 The above  forms allow us to  express the first member of a scale invariant field equation,
 which is thus a generalization of the first member of the field
equation of the General Relativity, including also scale invariance
as a fundamental property. This scale invariant first member is

\begin{equation}
R^{\nu}_{\mu}  \, - \, \frac{1}{2}  \, R \, g^{\nu}_{\mu} \, ,
\label{ff1}
\end {equation}

\noindent
with $R^{\nu}_{\mu} $ and $R$ given (\ref{RC}) and (\ref{RRR}). This first member depends on the $g^{\nu}_{\mu}$,
$\kappa_{\mu}$ and their derivatives. 
In General Relativity, the second member of the field equation writes
\begin{equation}
-  \, 8 \pi \, G  \,T'^{\nu}_{\mu} -\Lambda_{\mathrm{E}} \, g'^{\nu}_{\mu} \, .
\label{ff2}
\end{equation}

\noindent
The velocity of light $c$ is taken as  unity, $G$ is the gravitational constant, taken as an inscalar.
$T'^{\nu}_{\mu}$ is the energy-momentum tensor for a perfect fluid  in the system of General Relativity and
$\Lambda_{\mathrm{E}}$ is the cosmological constant of General Relativity, (we do not put a prime to
$\Lambda_{\mathrm{E}}$ , since the  index "E" is explicit enough).
The  second member of the scale invariant field equation must be an intensor,  as is the first one
given by (\ref{ff1}), it is   scale invariant  \citep{Canu77}. Thus, we have

\begin{equation}
 T'^{\nu}_{\mu} \, =  \,T^{\nu}_{\mu} \, ,
\label{T}
\end{equation}

\noindent
 where the right-hand  term is the  energy-momentum tensor in the new more general
coordinate system.
This has further implications, which are easily examined in the case of a perfect fluid \citep{Canu77}. We may write
(\ref{T}) as,

\begin{equation}
 ( p+\varrho) u_{\nu} u_{\mu} -g_{\nu \mu } p =
   ( p'+\varrho') u'_{\nu} u'_{\mu} -g'_{\nu \mu } p' \, .
\label{pr}
\end{equation}

\noindent
The velocities $u'^{\mu}$ and $u'_{\mu}$ transform like

\begin{eqnarray}
u'^{\mu}=\frac{dx^{\mu}}{ds'}=\lambda^{-1}  \frac{dx^{\mu}}{ds}=  \lambda^{-1} u^{\mu} \, ,   \; \\[2mm]
u'_{\mu}=g'_{\mu \nu} u'^{\nu}=\lambda^2 g_{\mu \nu} \lambda^{-1} u^{\nu} = \lambda \, u_{\mu} \, .
\end{eqnarray}

\noindent
Thus, relation(\ref{pr}) becomes  with (\ref{gmunu})

\begin{equation}
( p+\varrho) u_{\nu} u_{\mu} -g_{\nu \mu } p  =
 ( p'+\varrho') \lambda^2 u_{\nu} u_{\mu} - \lambda^2 g_{\nu \mu } p' \, .
\end{equation}

\noindent
This implies the following scaling of pressure and density in the new general coordinate system
\begin{equation}
 p =  p'  \, \lambda^2    \, \quad \mathrm{and} \quad   \varrho =  \varrho'  \, \lambda^2 \, .
\label{ro2}
\end{equation}

\noindent
Pressure and density are therefore not inscalars, but coscalars of power  -2.
For the empty space studied in this work, these relations do not intervene (since $p$ and $\varrho$ are zero),
but they   do  in the scale invariant cosmological equations where matter is present in the Universe.

Let us now consider the last term in   (\ref{ff2}),
 which contains $\Lambda_{\mathrm{E}}$ and is also globally scale invariant. Expression ({\ref{gmunu})
shows that the Einsteinian metric tensor  $g'_{\mu \nu}$ behaves
like $\lambda^2$, it is thus a cotensor of power +2. We can write  

\begin{equation}
\Lambda_{\mathrm{E}} \,  g'\,^\nu_\mu =  \Lambda_{\mathrm{E}}\, \lambda^2  \,  g^\nu_\mu   \, .
\label{ll}
\end{equation}
\noindent
We could possibly define a new $\Lambda$, by
\begin{equation}
\Lambda \, = \,  \Lambda_{\mathrm{E}} \, \lambda ^2 \, .
\label{L}
\end{equation}

\noindent
However, to avoid any ambiguity, we always keep all  expressions with $\Lambda_{\mathrm{E}}$, the true Einstein cosmological constant.
Thus, the second member of the scale invariant field equation becomes
\begin{equation}
-8 \pi G T^{\nu}_{\mu} - \lambda^2 \Lambda_{\mathrm{E}}  \, g^{\nu}_{\mu} \, .
\label{fff}
\end{equation}
\noindent
 With  (\ref{RC}), (\ref{RRR}),  (\ref{ff1}) and (\ref {fff}), the scale invariant field equation becomes \citep{Dirac73,Canu77}

\begin{eqnarray}
R'^{\nu}_{\mu}   -  \frac{1}{2}  \ g^{\nu}_{\mu} R' - \kappa^{; \nu}_{\mu}  - \kappa^{ \nu}_{;\mu} -2 \kappa_{\mu} \kappa^{\nu}
+ 2 g^{\nu}_{\mu}\kappa^{ \alpha}_{;\alpha} -g^{\nu}_{\mu}\kappa^{ \alpha} \kappa_{ \alpha} \nonumber \\   =
-8 \pi G T^{\nu}_{\mu} - \lambda^2 \Lambda_{\mathrm{E}}  \, g^{\nu}_{\mu} \, .
\label{field}
\end{eqnarray}

\noindent
The first member  only depends on
$g_{\mu \nu} $ and  $\kappa_\nu$ (or $\lambda$).  This   equation can be applied to various physical systems, 
characterized by their line element $ds^2$  and  their energy-momentum tensor $T^{\nu}_{\mu}$.
 Interestingly enough,  we shall see below  that
this equation, when applied  to the empty space, leads to useful relations between the $\kappa_\nu$ terms
and $\Lambda_{\mathrm{E}}$.

\section{Consequences of  scale  invariance of the empty space at
large scales}
 
\subsection{Relations between the scale factor $\lambda$ and the cosmological constant $\Lambda_{\mathrm{E}}$}

We consider the case of the empty  space,  with thus an energy-momentum
tensor $T^{\nu}_{\mu}$ equal to zero. The  line-element is given by the Minkowski metric,

\begin{equation}
ds^2 \, = \, c^2 dt^2 - (dx^2+dy^2+dz^2) \, .
\end{equation}
\noindent
 In General Relativity, the above metric implies that the first member of Einstein equation
is equal to zero,

\begin{equation}
R'^{\nu}_{\mu}  \, - \, \frac{1}{2}  \ g^{\nu}_{\mu} \, R'  \, = \, 0 \, .
\end{equation}
\noindent
Thus, in the scale invariant field equation (\ref{field}),  only  the following terms are remaining,

\begin{equation}
  \kappa^{; \nu}_{\mu}  + \kappa^{ \nu}_{;\mu} +2 \kappa_{\mu} \kappa^{\nu}
- 2 \, g^{\nu}_{\mu}\kappa^{ \alpha}_{;\alpha} +g^{\nu}_{\mu}\kappa^{ \alpha} \kappa_{ \alpha} = \, \lambda^2 \,  \Lambda_{\mathrm{E}} \, g^{\nu}_{\mu} \, .
\label{fcourt}
\end{equation}
\noindent
All other terms have disappeared and we are left only with a relation  between some functions of the scale factor $\lambda$
(through the $\kappa$-terms), the $g^{\nu}_{\mu}$  and the Einstein cosmological constant. 
The term $\kappa_{\nu}$ is related to   $\lambda$ and on its first two derivatives according to relation (\ref{kappa}).
It is interesting to  remark that the cosmological constant which can be interpreted as
 the energy density of the vacuum is also related,
 in the present context, to the properties of scale invariance in the empty space.

At this stage, it may be opportune to
recall that the problem of the cosmological constant in the empty space is not a new one.  \citet{Bert90} are  quoting  the following remark
they got from Professor  Bondi, who stated that :  {\it{ " Einstein's disenchantment with the cosmological constant was partially motivated by a desire to preserve scale invariance
of the empty space Einstein equations ".}} This remark is in agreement with the fact that $\Lambda_{\mathrm{E}}$ is not scale invariant 
as are the $T^{\nu}_{\mu}$. 
  The above developments show that the  scale invariant theory may offer a possibility
to conciliate  the existence of $\Lambda_{\mathrm{E}}$ with the scale invariance of the empty space. This reconciliation takes the form
 of relations (\ref{fcourt}), which are now further analyzed.

As stated above, the scale factor $\lambda$ can only be  a function of time,  therefore only the zero component
of $\kappa_\mu$ is non-vanishing. Thus, the coefficient of metrical connection $\kappa^{; \nu }_\mu$ is

\begin{eqnarray}
\kappa^{; \nu }_\mu = \kappa_{\mu;\rho} g^{\rho \nu}= \kappa_{0;0} g^{0,0}= \partial_0 \kappa_0 =  \frac{d \kappa_0}{dt} \equiv
{\dot{\kappa}_0} \, .
\end{eqnarray}
\noindent
The 0 and   the 1, 2, 3 components  of what remains from the field equation (\ref{fcourt})  become respectively

\begin{equation}
3 \kappa^2_0 \, = \,\lambda^2 \, \Lambda_{\mathrm{E}},
\label{k1}
\end{equation}
\begin{equation}  
 2  \dot{\kappa}_0 - \kappa_0^2 = -\lambda^2  \Lambda_{\mathrm{E}}   \, .
\label{k2}
\end{equation}
\noindent
From the definition (\ref{kappa}), one has $\kappa_0 \, = - \dot{\lambda}/\lambda$ (with c=1 at the denominator)
and expressions (\ref{k1}) and (\ref{k2}) lead to   the two following differential relations for $\lambda$,

\begin{eqnarray}
\  3 \, \frac{ \dot{\lambda}^2}{\lambda^2} \, =\, \lambda^2 \,\Lambda_{\mathrm{E}}  \,  \label{diff1}  \\ [2mm]
 \quad \mathrm{and} \quad 2 \, \frac{\ddot{\lambda}}{\lambda} - \frac{ \dot{\lambda}^2}{\lambda^2} \, =
\, \lambda^2 \,\Lambda_{\mathrm{E}}  \, .
\label{diff2}
\end{eqnarray}
\noindent
These expressions   may also be written in equivalent forms

\begin{eqnarray}
\frac{\ddot{\lambda}}{\lambda} \, = \,  2 \, \frac{ \dot{\lambda}^2}{\lambda^2} \, , \label{diff3} \\  [2mm]
 \quad \mathrm{and} \quad \frac{\ddot{\lambda}}{\lambda} -\frac{ \dot{\lambda}^2}{\lambda^2} \, = \, \frac{\lambda^2 \,\Lambda_{\mathrm{E}}}{3} \, ,
\label{diff4}
\end{eqnarray}
\noindent
These are  the relations between Einstein's cosmological constant $\Lambda_{\mathrm{E}}$ and the scale factor $\lambda$.
Relation (\ref{diff3}), that does not contain $\Lambda_{\mathrm{E}}$, also places a constraint on $\lambda(t)$.
These various  relations may intervene in  the equations of cosmology and, as a matter of fact, they will reveal themselves most useful
in leading to valuable simplifications.

Let us examine the solution of the  differential equation (\ref{diff3}).
We  consider a solution of the form
\begin{equation}
\lambda \,= \, a\,  (t \,- \,b)^n+d \, .
\label{bl}
\end{equation}
\noindent
 Equation (\ref{diff3}) imposes
$d \,= \,0$, meaning there is no additive constant to $\lambda$. It also implies $n \, = \, -1$. Interestingly enough,  there is no
constraint on the value of $b$, which means that the  origin $b$ of the timescale is not determined by the above equations.
 (The origin of the time, in the scale invariant cosmology like in other cosmologies, will be determined by 
 the solutions of the equations of the particular cosmological model considered).
The constant $a$ can be fixed for example by (\ref{diff4}), which gives 

\begin{equation}
a \, = \, \sqrt{\frac{3}{\Lambda_{\mathrm{E}}}} \,  .
\end{equation}
\noindent
In physical units, we would have $a \, = \, \sqrt{3/(c^2 \, \Lambda_{\mathrm{E}}})$. Thus, we may  finally write $\lambda$ as follows,

\begin{equation}
\lambda \, = \, \sqrt{\frac{3}{\Lambda_{\mathrm{E}}}} \, \frac{1}{c \,t}  \, .
\label{lamb}
\end{equation}
\noindent
We see that the scale invariance of the empty space at macroscopic and large scales imposes a scale factor $\lambda$, such that 
$\lambda^2$ is related to the inverse of the energy density of the vacuum (to which  $\Lambda_{\mathrm{E}}$ is proportional).
The factor $\lambda$ varies  like the inverse  of the cosmic time $t$, with no origin fixed yet at this stage of the developments.

\subsection{Further remarks on the scale factor $\lambda$}

The scale invariant equations
are identical to those of  General Relativity   at a given fixed time, which we may choose to be the present one $t_0$.
 Some  departures from  General Relativity may appear when  the evolution of a physical effect over the ages is intervening. Then, 
there may be different values of $\lambda$ at different epochs.
At this stage, it is difficult to anticipate about the kind and size of the effects resulting from
 scale invariance, but we
will see in future works that the main effect is a cosmic acceleration of comoving galaxies.

The fact that $\Lambda_{\mathrm{E}}$ and  the energy of the vacuum are  related in General Relativity implies,
since $\lambda$, its derivatives  and  $\Lambda_{\mathrm{E}}$ are  connected, that there may also be an energy-density associated 
the scale factor and its variations. The exact form of this energy-density will be studied on the basis of the appropriate cosmological
equations.

Noticeably in the present framework, if the Einstein cosmological constant $\Lambda_{\mathrm{E}}$ is different  from zero, the scale factor 
$\lambda$ must necessarily vary with time.
Reciprocally,  if $\lambda$ is a constant, the cosmological constant should be zero.
 This conclusion  depends on the
assumption that the vacuum at large scales is scale invariant.

An estimate of $\Lambda_{\mathrm{E}}$ can be obtained from (\ref{lamb}), if we assume for example that $\lambda=1$ 
at the present time $t_0$. We get in physical units
\begin{equation}
  \Lambda_{\mathrm{E}} \, = \, \frac{3}{c^2 \, t^2 _0} \, .
\label{LE}
\end{equation}
\noindent
For an age of the Universe of 13.8 Gyr \citep{Frie08}, we obtain $\Lambda_{\mathrm{E}}= 1.59 \cdot 10^{-35} \; \mathrm{s}^{-2}$  (with c=1)  or in current physical units
$\Lambda_{\mathrm{E}}= 1.76 \cdot 10^{-56} \; \mathrm{cm}^{-2}$.  Expression (\ref{LE}) is similar
to classical expressions  and it  gives a numerical  estimate in agreement with  the  current value (see for example \citep{Carm01})
derived from the observations of cosmic acceleration.

\section{Conclusions}

We have made  the hypothesis that the empty space is scale invariant at large  scales,
 where General Relativity also applies. This hypothesis  has  far reaching consequences.
 It allows, referring to the above mentioned  Bondi's remark 
(Sect. 2.1), to keep a cosmological constant and preserve at the same time the scale invariance of the empty space. It also
imposes some relations between the cosmological constant $\Lambda_{\mathrm{E}}$ and the relative variations 
of the scale factor $\lambda$ that
describes how the space-time intervals behave as a function of coordinates. In a framework consistent with the Cosmological Principle of homogeneity
and isotropy, the scale factor $\lambda$  only depends on time $t$.


Finally, we note that it may not be surprising that the assumption of the  scale invariance of the  empty space at large scales
 leads to some relations
involving the  cosmological constant, since this  constant   represents the energy density of the vacuum in General Relativity.
The above challenging hypothesis needs to be further incorporated in consistent cosmological models, which
can be rejected or supported  only by a careful  analysis of its  observational consequences.

\vspace*{5mm}

\noindent
Acknowledgments: I want to express my best thanks to the physicist D. Gachet and Prof. G. Meynet for their 
continuous encouragements.

\bibliographystyle{aa}
\bibliography{Maeder-article-I}
\end{document}